\documentstyle[preprint,aps]{revtex}
\begin{document}
\draft

\title{Interface Transparency of Nb/Pd Layered Systems}
\author{C. Cirillo,
S.L. Prischepa,\footnote{Permanent address: State University of
Computer Science and RadioElectronics, P. Brovka street 6, 220600,
Minsk, Belarus.} M. Salvato and C. Attanasio\footnote
{Corresponding author. Tel.: +39-089-965288; Fax: +39-089-965275;
E-mail: attanasio@sa.infn.it} }
\address{Dipartimento di Fisica ``E.R. Caianiello" and INFM, Universit\`{a} degli
Studi di Salerno, Baronissi (Sa), I--84081, Italy}
\date{\today}
\maketitle

\begin{abstract}
We have investigated, in the framework of proximity effect theory,
the interface transparency $\cal{T}$ of superconducting/normal
metal layered systems which consist of Nb and high paramagnetic Pd
deposited by dc magnetron sputtering. The obtained $\cal{T}$ value
is relatively high, as expected by theoretical arguments. This
leads to a large value of the ratio $d_{s}^{cr}/ \xi_{s}$ although
Pd does not exhibit any magnetic ordering.
\end{abstract}


\section{Introduction}
Interface transparency $\cal{T}$ of artificial layered systems is
an interesting issue of study, both for its fundamental and
practical consequences and many papers have been recently devoted
to this topic \cite{Aarts,Rusanov,Otop,Attanasio}. From one side,
in fact, $\cal{T}$ is related to differences between Fermi
velocities and band-structures of the two metals. On the other
hand it is an essential parameter to take into account in the
study of depairing currents \cite{Geers} and quasiparticle
injection devices \cite{Vasko,Dong,Yeh,Sefrioui} where high
interface transparency is an important ingredient.

In this article we performed a proximity effect study of Nb/Pd
layered system \cite{Cirillo} taking into account the essential
ingredient of interface transparency. We chose Nb as
superconducting material and Pd as normal metal. The choice of Nb
was related to its highest critical temperature among the
superconducting elements while, among normal metals, Pd is the one
with the larger spin susceptibility \cite{Nieuwenhuys} which leads
to giant magnetic moments in some dilute Pd alloys \cite{Dam}.
Moreover, by theoretical argument based on Fermi velocities and
band-structures mismatch, we expected a more transparent interface
than in other superconductor/normal metal combinations, such as,
for example, Nb/Cu \cite{PhD}.

\section{Theoretical background}
When a superconductor (S) comes into contact with another material
(X) proximity effect occurs. The other material can be a
superconductor with a lower transition temperature (S'), a normal
metal (N), a ferromagnet (F) or a spin glass (M). In any case
there is a mutual influence which depresses superconductivity in S
and induces superconductivity in X. Since at the interface the
order parameter decreases in S over the coherence length
$\xi_{s}$, it is necessary a minimum thickness of the S layer,
$d_{s}$, to make superconductivity appear. If $d_{s}$ is small the
order parameter cannot reach its maximum value and the critical
temperature $T_{c}$ of the system is reduced, until $d_{s}$
becomes too thin and superconductivity is lost. The thickness at
which it happens is called critical thickness, $d_{s}^{cr}$.   On
the other hand, Cooper pairs coming from S penetrate X, but they
are broken up over a characteristic length $\xi_{x}$, depending on
the pair breaking mechanism in X. At finite temperature pairs
loose their phase coherence by thermal fluctuations: this is the
only pair breaking mechanism present in N metals, and lead to a
temperature dependent characteristic distance, $\xi_{n}(T)$, which
can become large at low temperatures. In magnetic metals pair
breaking is due to the exchange interaction $E_{ex}$ which acts on
the spin of the Cooper pairs. For strong magnets, such as Fe,
$E_{ex}>>k_{B}T$: this leads to a few Angstrom temperature
independent coherence length $\xi_{F}$ in the magnetic layer
\cite{Aarts,Koorevaar,Verbanck}.

Anyway interfaces between different metals are never fully
transparent with the result that proximity effect is somehow
screened, because electrons coming from S are reflected rather
than transmitted in X. A finite transparency gives rise, for
example, to a smaller $d_{s}^{cr}$. This may be due to interface
imperfections, lattice mismatches, fabrication method
\cite{Attanasio,Schock}, but also to intrinsic effects such as
difference between Fermi velocities and band-structures of the two
metals \cite{PhD}. The interface transparency due to Fermi
velocities mismatches in the free electron model, is given by
\cite{Aarts,Golubov}:

\begin{equation}
{\cal{T}} = \frac{4 k_{x} k_{s}}{[k_{x} +
k_{s}]^{2}}\label{eq:Tth}
\end{equation}

\noindent where $k_{x,s}= {m v_{x,s}}/{\hbar}$ are the projections
of Fermi wave vectors of X and S metals on the direction
perpendicular to the interface. Moreover for the magnetic case the
situation is more complicated due to the role played by the
splitting of the spin subbands and the spin-dependent impurity
scattering \cite{deJong}.

The starting point for a complete description of proximity effect
in multilayers, valid for arbitrary transparency, was given by
Kupriyanov and Lukichev \cite{Kupriyanov} in the framework of
Usadel equations (dirty limit). In particular, the model we used
to describe the dependence $T_{c}(d_{s})$ for N/S/N trilayers is
based on the Werthamer approximation, valid for not too low
temperatures, provided the boundary transparency is sufficiently
small \cite{Golubov}. In this limit the system of algebraic
equations to determine $T_{c}$ is:

\begin{equation}
\Omega_{1}\tan\Bigg(\frac{\Omega_{1} d_{S}}{2 \xi_{S}}\Bigg) =
\frac{\gamma}{\gamma_{b}} \label{eq:gol1}
\end{equation}

\begin{equation}
\Psi \Bigg(\frac{1}{2} + \frac{\Omega_{1}^{2} T_{cs}}{2
T_{c}}\Bigg) - \Psi \Bigg(\frac{1}{2}\Bigg) = \ln \Psi
\Bigg(\frac{T_{cs}}{T_{c}}\Bigg),
\frac{T_{cs}}{T_{c}}\gg\frac{\gamma}{\gamma_{b}} \label{eq:gol2}
\end{equation}

\noindent with the identification of the Abrikosov-Gorkov
pair-breaking parameter $\rho$=$\pi T_{c} \Omega_{1}^{2} =\pi
T_{c} (\gamma/\gamma_{b}) (2 \xi_{s}/d_{s})$ where $\Psi(x)$ is
the digamma function and $T_{cs}$ is the bulk critical temperature
of the S layer. These equations contain two parameters $\gamma$
and $\gamma_{b}$ defined as

\begin{equation}
\gamma = \frac{\rho_{s} \xi_{s}}{\rho_{n} \xi_{n}},
\gamma_{b} = \frac{R_{B}}{\rho_{n} \xi_{n}} \label{eq:gamma}
\end{equation}

\noindent where $\rho_{s}$ and $\rho_{n}$ are the low temperature
resistivities of S and N, respectively, while $R_B$ is the
normal-state boundary resistivity times its area. The parameter
$\gamma$ is a measure of the strength of the proximity effect
between the S and N metals and can be determined experimentally by
measuring $\rho_s \equiv \rho_{Nb}$, $\rho_n \equiv \rho_{Pd}$,
$\xi_s \equiv \xi_{Nb}$ and $\xi_n \equiv \xi_{Pd}$. The parameter
$\gamma_{b}$, instead, describes the effect of the boundary
transparency $\cal{T}$, to which it is roughly related by

\begin{equation}
{\cal{T}} = \frac{1}{1 + \gamma_{b}} \label{eq:T}
\end{equation}

\noindent Due to its dependence on $R_B$, which is difficult to
measure, $\gamma_b$ (or $\cal{T}$) can't be determined
experimentally, so it was extracted by a fitting procedure.

\section{Experimental results}

The samples were grown on Si(100) substrates by a dual-source
magnetically enhanced dc triode sputtering system and they consist
of Nb layers ($T_c \approx 8.8$ K) and Pd layers. The deposition
conditions were similar to those of the Nb/Pd multilayers earlier
described \cite{Cirillo} except for the fact that the 8 samples,
obtained in a single deposition run, were not heated. Three
different sets of multilayers were prepared. Two sets (set A and
set B), built as follows, $d_{Pd}/d_{Nb}/d_{Pd}$, were used to
determine $d_{s}^{cr} \equiv d_{Nb}^{cr}$ by the variation of
$T_c$ as function of the Nb layer thickness. Here $d_{Pd}$ was
fixed at around 1500 \AA\  in order to represent a half-infinite
layer, while $d_{Nb}$ was variable from 200 to 1300 \AA. The third
set (set C) was used to estimate $\xi_{Pd}$ by the variation of
$T_c$ with $d_{Pd}$. Now the samples were made up of five layers:
$d_{Pd}^{out}$/$d_{Nb}$/$d_{Pd}^{in}$/$d_{Nb}$/$d_{Pd}^{out}$,
with the outer Pd layers of 300 \AA\ in order to create a
symmetric situation for the Nb layers, with $d_{Nb}$ fixed at 500
\AA, while $d_{Pd}^{in}$ was varied from 50 to 300 \AA. Extensive
low and high angle X-ray diffraction patterns has been performed
to structurally characterize the samples. High angle scans clearly
showed the Nb(110) and the Pd(111) preferred orientations and
allow us to estimate the lattice parameters, $a_{Nb}=3.3$ \AA\ for
the bcc-Nb and $a_{Pd}=3.9$ \AA\ for the fcc-Pd, in agreement with
the values reported in literature \cite{Kittel}. Low angle
reflectivity measurements on samples deliberately fabricated to
perform structural characterization, show a typical interfacial
roughness of 12 \AA\ \cite{Cirillo}.

The superconducting properties, transition temperatures $T_c$,
perpendicular and upper critical magnetic fields $H_{c2\bot}(T)$
and $H_{c2 \|}(T)$ were resistively measured using a standard dc
four-probe technique. The values obtained for the resistivities
are independent of the layering and in the range 3-4 $\mu \Omega
\cdot cm$ at 10 K. The ratios $\rho_{N}$(T=300
\,K)/$\rho_{N}$(T=10 \,K), with $\rho_{N}$ the normal state
resistivity, were in the range 1.7-2.2 for all the series
confirming the high uniformity of the transport properties in the
samples obtained in the same deposition run. Measuring a
resistivity value of $\rho_{Nb}$=2.5 $\mu \Omega \cdot cm$ in the
case of a deliberately fabricated 1000 \AA\ thick single Nb film,
and assuming a parallel resistor model \cite{Gurvitch}, we deduced
$\rho_{Pd}\approx 5 \mu \Omega \cdot cm$.

\section{Results and discussion}

In Fig. 1 the critical temperature $T_c$ is reported as a function
of $d_{Nb}$ for the $d_{Pd}$/$d_{Nb}$/$d_{Pd}$ trilayers . The
transition temperature of the sample with $d_{Nb}=200$ \AA\ is not
reported since it was below 1.75 K, the lowest temperature
reachable with our experimental setup. The temperature asymptotic
value of 8.8 K for our bulk Nb is reached above 1500 \AA\ while,
below 450 \AA, $T_c$ is sensitively reduced. Moreover in Fig. 1
the transition temperatures $T_{c}(d_{Nb})$ are compared to those
of single Nb films, clearly indicating that the suppression of the
critical temperatures of the trilayers comes indeed from the
proximity effect rather than from the $T_{c}$ thickness dependence
of single Nb. Fig. 2 shows $T_c$ vs $d_{Pd}^{in}$ measurements
performed on the
$d_{Pd}^{out}$/$d_{Nb}$/$d_{Pd}^{in}$/$d_{Nb}$/$d_{Pd}^{out}$
systems. With increasing $d_{Pd}^{in}$ the critical temperature is
lowered until a non monotonic behaviour with a minimum for
$d_{Pd}^{in} \approx 140$ \AA\ is reached, then the curve levels
off to a value of 7.8 K for large $d_{Pd}^{in}$. A similar
behavior, consisting in a dip before reaching the maximum and then
the asymptotic value, was found in S/F systems such as $\rm V/Fe$,
$\rm V/Fe_xV_{1-x}$ \cite{Aarts}, $\rm Nb/Pd_{1-x}Fe_x/Nb$
\cite{Schock} and $\rm Nb/Cu_{1-x}Ni_x$ \cite{Rusanov,Boogaard},
and may be related to the strong paramagnetic nature of Pd, such
as the abrupt decrease of $T_c$ for small values of $d_{Pd}^{in}$.
A qualitatively explaination for the saturation of the
$Tc(d_{Pd}^{in})$ curve can be given by considering that when the
Nb layers are separated by a thin Pd layer, the decay of Cooper
pairs from both sides overlap, the $T_c$ of the system is
increased and we say that the Nb layers are coupled. By increasing
$d_{Pd}^{in}$ the Nb layers become more and more decoupled and the
critical temperature reaches a limiting value related to $T_c$ of
the single isolated Nb layer ($d_{Nb}=500$ \AA). We found that
this critical temperature value is a little higher than the one
obtained for the trilayer with $d_{Nb}=500$ \AA\ ($T_c \approx 6$
K, see Fig. 1). This is probably due to different deposition
conditions, since these two series were fabricated in different
deposition runs. Moreover, in the two systems, Nb layers were
included in Pd layers of different thickness and this may also
play a role. The thickness for which the temperature becomes
constant is the decoupling thickness $d_{Pd}^{dc}$. This thickness
is related to the coherence length by $d_{Pd}^{dc}\approx$ 2
$\xi_{Pd}$. We identify $d_{Pd}^{dc}$ extrapolating the steepest
slope in the $T_c$($d_{Pd}^{in}$) curve to the $d_{Pd}^{in}$ axis
(line in Fig. 2) \cite{Geers}. The value for $\xi_{Pd}$ of
approximately 60 \AA\ that we find with this procedure is
comparable with other values reported in literature for similar
systems \cite{Schock} while it is considerably lower than the
values found for other normal metals, such as Cu
\cite{Geers,Boogaard}, and greater than values found for the
ferromagnetic ones \cite{Aarts,Dam,Koorevaar}. In addition this
value, in our temperature range, is in agreement with the one
estimated from the measured $\rho_{Pd}$ with the expression of
$\xi_{Pd}$ valid in the dirty limit:

\begin{equation}
\xi_{Pd}= \sqrt \frac{\hbar D_{Pd}}{2 \pi k_B T}. \label{eq:csipd}
\end{equation}

\noindent Here $D_{Pd}$ is the diffusion coefficient which is
related to the low temperature resistivity $\rho_{Pd}$ through the
electronic mean free path $l_{Pd}$ by \cite{Broussard}

\begin{equation}
D_{Pd}=\frac{v_{Pd}l_{Pd}}{3}\label{eq:Dpd}
\end{equation}

\noindent in which

\begin{equation}
l_{Pd}=\frac{1}{v_{Pd}\gamma_{Pd}\rho_{Pd}}\Bigg(\frac{\pi
k_{B}}{e}\Bigg)^{2}\label{eq:lpd}
\end{equation}

\noindent where $\gamma_{Pd}\approx11.2\times10^{2} J/K^{2}m^{3}$
is the Pd electronic specific heat coefficient \cite{Handbook} and
$v_{Pd}$=2.00 $\times$ 10$^7$ $cm/s$ is the Pd Fermi velocity
\cite{vfermipd}. The values obtained for $\xi_{Pd}$ are between 73
\AA\ and 115 \AA\ for T=10 K and T=4 K, respectively, while, from
Eq. (\ref{eq:lpd}), $l_{Pd}$=$60$ \AA. The value of the ratio
$l_{Pd}/\xi_{Pd}$, always less than one in the considered
temperature range, confirms the validity of the dirty limit
approximation.

Inspired by these results we also tried to explain the abrupt
decrease and the dip of $T_c$ shown in Figure 2 extending the
Radovic theory \cite{Radovic} to the case of S/N systems with N a
normal metal with high spin susceptibility. While Radovic's theory
well describes this behaviour for both S/F
\cite{Koorevaar,Verbanck} and S/M systems \cite{Carmine}, we did
not obtain a good agreement with the experimental data. However we
have to remark that for Nb/Pd multilayers a monotonic decrease of
$T_{c}$($d_{Pd}$) was observed \cite{Cirillo,Schock,Kaneko}. This
behaviour has been discussed in the framework of de
Gennes-Werthamer theory, but at the price of supposing very low
or, alternatively, very high values of Pd resistivities
\cite{Cirillo,Kaneko}.

Upper critical fields of $d_{Pd}$/$d_{Nb}$/$d_{Pd}$ trilayers were
also measured in order to determine $\xi(0)$, the Ginzburg-Landau
coherence length at zero temperature, from the slope
$S=dH_{C2}/dT|_{T=T_{c}}$. $\xi_{Nb}$ is, in fact, related to
$\xi(0)$ by the relation $\xi_{Nb}$=2$\xi(0)$/$\pi$. The slope $S$
is extrapolated from the $H_{C2\bot}(T)$ curves by a linear fit
near $T_{c}$, as shown in Fig. 3. A value of about $\xi_{Nb}$=64
\AA\ was found for single Nb film, $1000$ \AA\ thick and this
value agrees with the one obtained for samples with thicker Nb
interlayers ($d_{Nb} \geq 700$ \AA). It is also interesting to
note that $H_{C2\bot}(T)$ slopes for the different samples are
quite constant, which is a behaviour already observed in S/F
systems \cite{Koorevaar}. In Fig. 4 parallel critical magnetic
fields are shown. The main feature of these curves is the square
root behaviour of $H_{C2\|}(T)$ in all the temperature range and
the absence of the $3D-2D$ crossover. Also this feature may be
related to the magnetic nature of Pd. So, from both perpendicular
and parallel critical fields measurements seems to emerge that
Nb/Pd systems behave more as a S/F rather than a S/N system, even
if this indication is not confirmed by $T_{c}$'s measurements. A
similar behaviour was alrealy observed in Nb/Pd multilayers
\cite{Cirillo}. In this case the critical temperature showed a
monotonic decrease as a function of the Pd thickness, which was
described in the framework of the classical Gennes-Werthamer
proximity theory valid for S/N systems. On the other hand the
hypothesis of the Pd ferromagnetic nature seemed to be the reason
of the early $3D-2D$ dimensional crossover observed in
$H_{C2\|}(T)$ measurements \cite{Cirillo}.

With these results for $\xi_{Nb}$ and $\xi_{Pd}$ and with the
measured $\rho_{Nb}$ and $\rho_{Pd}$ values reported above it is
possible to calculate the proximity effect parameter $\gamma$=0.53
and to reproduce $T_{c}$($d_{Nb}$) of the trilayers by Eq.
(\ref{eq:gol2}) with only one free parameter, $\gamma_{b}$. As
reported above in Eq. (\ref{eq:gol2}), the validity regime of the
Werthamer approximation is $T_{c}/T_{cs}\gg\gamma/\gamma_{b}$. In
our case, even if the condition is not fully satisfied, the ratio
$\gamma/\gamma_{b}=0.4$ is always less than $T_{c}/T_{cs}$. In
fact $T_{c}/T_{cs}$, in the trilayers critical temperatures range,
is between $0.5-1$. The result of the calculation is shown in Fig.
1 (solid line) and it is obtained for $\gamma_{b}$=1.17, which
means $\cal{T}$=0.46. In Fig. 1 are also shown the curves obtained
for different $\cal{T}$ values ($\cal{T}$=0.42, 0.46, 0.54 from
left to right). It is evident that varying $\cal{T}$ only of
$0.04$ the accordance between the theory and the experimental data
is completely lost. In addition, motivated by the observed
$H_{C2\bot}(T)$ and $H_{C2\|}(T)$ behaviours, we tried to
reproduce the experimental results with the extended theory for
$S/F$ systems \cite{Aarts}. However in this case the best fit to
the data is obtained for $\cal{T}$=0.86, which seems to be an
unphysical value for the transparency of a real system. Of course
if we also identify $\xi_{Pd}\approx$ $d_{Pd}^{dc}$, as recently
reported \cite{Tagirov,Fominov}, the interface transparency will
be further increased. In particular we are not able to reproduce
the experimental point even if we use $\cal{T}$=0.99. From the
curve in Fig. 1 it is possible to determine the value of
$d_{Nb}^{cr}\approx 200$ \AA. In Table I are summarized all the
samples parameters. The critical thickness normalized to the
coherence length in Nb can also be calculated. It depends both on
interface transparency and on the strength of the pairing,
lowering with increasing $\xi_{Pd}$ and decreasing $\cal{T}$. In
S/F systems with $\cal{T}$=1, $d_{s}^{cr}$/$\xi_{s}$ has its
theoretical upper limit close to 6 \cite{Aarts}. The value we
obtained, $d_{Nb}^{cr}$/$\xi_{Nb}\approx 3$, is comparable to that
($d_{Nb}^{cr}$/$\xi_{Nb}=2.69$) reported for $\rm
Nb/Cu_{0.915}Mn_{0.085}$ systems having $\cal{T}$=0.33
\cite{Geers} and sensitively higher than the one found for Nb/Cu
($d_{Nb}^{cr}$/$\xi_{Nb}=0.48$) with $\cal{T}$=0.29 \cite{Geers}.
The ratio is lower than ours also for $\rm Nb/Cu_{1-x}Ni_{x}$,
probably due to the little interface transparency of those systems
\cite{Rusanov,Boogaard}. The obtained $\cal{T}$ value for Nb/Pd
systems is substantially high, although lower than expected on
theoretical argument based on Fermi velocities mismatch. The
values $v_s \equiv v_{Nb}$=2.73 $\times$ 10$^7$ $cm/s$
\cite{vferminb} and $v_n \equiv v_{Pd}$=2.00 $\times$ 10$^7$
$cm/s$ \cite{vfermipd} in Eq. (\ref{eq:Tth}), in fact, would yield
$\cal{T}$=0.98. This happens also for other S/N systems, such as
Nb/Cu, where ${\cal{T}}_{exp}\approx 0.29-0.33$
\cite{Geers,Boogaard}, while $\cal{T}$=0.8 should be expected, for
Nb/Al with ${\cal{T}}_{exp}\approx 0.2$-0.25 \cite{Geers,NbAl}
instead of $\cal{T}$=0.8, and for Ta/Al with
${\cal{T}}_{exp}\approx 0.25$ \cite{Geers}. Another very important
factor to have a high $\cal{T}$ value is the matching between
band-structures of the two metals. This parameter influences
transparency even more than Fermi velocities. Conduction electrons
in Nb and Pd have both a strong d-character
\cite{Geers,vfermipd,Nieuwenhuys} and this fact would also lead to
high values of $\cal{T}$. Anyway, as we said, also lattice
mismatches play a role. In Nb/Pd systems the growth of the bcc-Nb
structure on the fcc-Pd one may induce stress at the interfaces.
High values of the interface roughness reported in literature for
similar Nb/Pd systems fabricated with different deposition
techniques \cite{Cirillo,Schock,Kaneko} are consistent with these
considerations. In this sense a systematic study of the influence
of the fabrication method on interface transparency, as already
done on Nb/Cu systems \cite{Attanasio}, would be interesting.
Since preparations methods seem to have a strong influence on
$\cal{T}$, we could expect to have samples of higher quality, and
consequently, of larger transparency using, for example, Molecular
Beam Epitaxy (MBE) deposition technique.

\section{Conclusions}

In conclusion we have studied the proximity effect in Nb/Pd
layered systems using the interface transparency as the only free
parameter. We obtained a relatively high value for $\cal{T}$,
higher than the one found for Nb/Cu, in accordance with
theoretical considerations about mismatches between Fermi
velocities and band-structures of the two metals. Anyway the value
obtained for $\cal{T}$ can only be indicative: it depends on
several factors, such as the way we extracted $\xi_{Pd}$ from the
$T_{c}(d_{Pd}^{in})$ curves neglecting its temperature dependence,
the measured values of $\rho_{Nb}$ and $\rho_{Pd}$ and the
approximation used to go from $\gamma_{b}$ to $\cal{T}$. What
emerges from this study is that Nb/Pd is, in a sense, an
intermediate system between the well known Nb/Cu and other S/F or
S/M systems such as Nb/Fe, V/Fe, $\rm Nb/Cu_{1-x}Ni_x$ or Nb/CuMn.
The high values of the ratio $d_{Nb}^{cr}/\xi_{Nb}$ and the lack
of agreement between Radovic's theory and experimental results can
be explained by good interface transparency rather than by
magnetic arguments. In this sense it is useful to compare our
result with the one obtained for Nb/CuMn. The ratio
$d_{s}^{cr}$/$\xi_{s}$ is comparable for the two systems
\cite{PhD}. On the other hand the stronger magnetic nature of CuMn
is known and also indirectly demonstrated by Radovic's fit, that
well describes the critical temperature behavior for Nb/CuMn
multilayers \cite{Carmine} but not for Nb/Pd. This makes us think
that Pd-based magnetic alloys are good candidates for studying the
S/F proximity problem: very low impurity concentrations will
induce ferromagnetic ordering, but should not produce great
disorder at the interface. An interesting alloy could be PdNi: the
magnetic order starts to appear for a Ni concentration of 2.5 \%.
This makes the alloy stoichiometry easy to control and induces an
homogeneous ferromagnetism, with a relatively low Curie
temperature \cite{Kontos1,Kontos2}. In this sense PdNi seems to be
more intriguing than CuNi because of the low values of the
interface transparency measured in Nb/CuNi systems.

%\end{document}
\begin{table}
\caption{Values of the electrical resistivities $\rho_{Nb}$ and
$\rho_{Pd}$, of the coherence lengths $\xi_{Nb}$ and $\xi_{Pd}$ as
experimentally determined and used in the fit procedure to
estimate the ratio $d_{Nb}^{cr}$/$\xi_{Nb}$, and
the transparency
parameter $\cal{T}$.} \label{table}
\begin{tabular}{cccccc}
\rm $\rho_{Nb}$ $(\mu \Omega \cdot cm)$ & $\rho_{Pd}$ $(\mu \Omega
\cdot cm)$ & $\xi_{Nb}$ $(\rm \AA)$ & $\xi_{Pd}$ $(\rm \AA)$ &
$d_{Nb}^{cr}$/$\xi_{Nb}$ & $\cal{T}$\\
\hline
2.5 & 5.0 & 64 & 60 & 3.2 & 0.46  \\
\end{tabular}
\end{table}

\begin{figure*}
\caption{Critical temperature $T_c$ versus Nb thickness $d_{Nb}$
for $d_{Pd}$/$d_{Nb}$/$d_{Pd}$ trilayers (solid symbols).
Different symbols refer to sample sets obtained in different
deposition runs (set A and set B). Open symbols refer to single Nb
films. The dashed line shows the $T_c$ of our bulk Nb. The solid
line is the result of the calculations with the parameters given
in Table I. The arrow indicates the value of $d_{Nb}^{cr}$. The
dashed and the dot-dashed lines indicate the theoretical
calculation for $\cal{T}$=0.42 and $\cal{T}$=0.54, respectively. }
\vspace{0.5in} \caption{Critical temperature $T_c$ versus inner Pd
thickness $d_{Pd}^{in}$ for sample set C. The arrow shows the
value of $d_{Pd}^{dc}$. The line indicates the method used to
determinate it.} \vspace{0.5in} \caption{Perpendicular critical
magnetic field for $d_{Pd}$/$d_{Nb}$/$d_{Pd}$ trilayers and for a
single Nb film. Different symbols represent different $d_{Nb}$, as
indicated in the legend. The lines are the result of the linear
fits near $T_c$.} \vspace{0.5in} \caption{Parallel critical
magnetic field for the same $d_{Pd}$/$d_{Nb}$/$d_{Pd}$ trilayers
with different $d_{Nb}$, as in Fig. 3.} \vspace{0.5in}
\end{figure*}

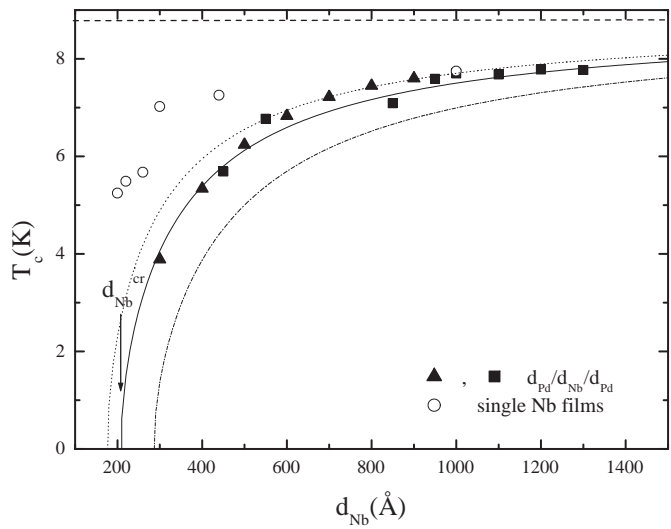
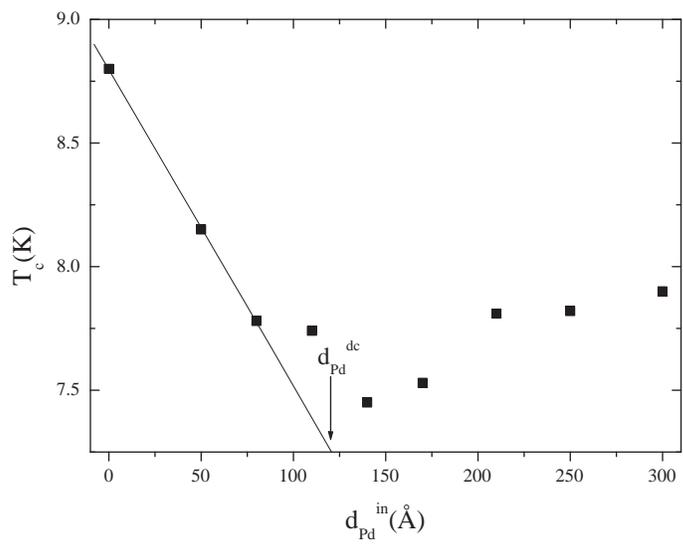
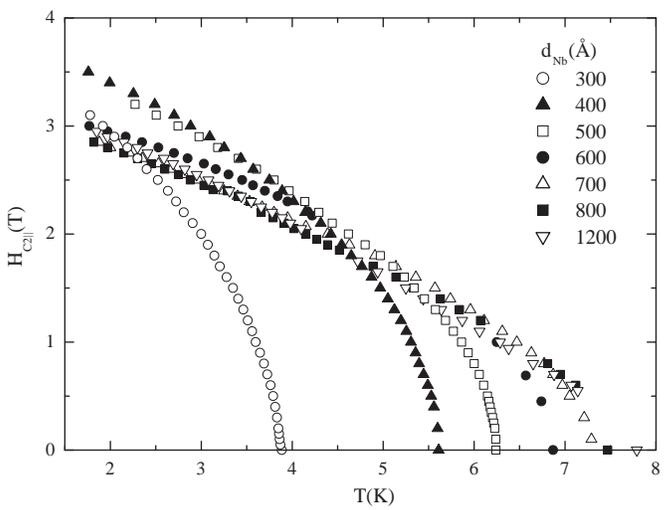
\end{document}